\documentclass[aps,prl,twocolumn,showpacs,10pt,nofootinbib]{revtex4}
\usepackage[dvips]{graphicx}
\usepackage{amsmath}
\usepackage{amssymb}
\usepackage{mathrsfs}
\usepackage{ulem}
\usepackage[mathscr]{euscript}

\begin{document}

\title{T-Carbon: A Novel Carbon Allotrope}
\author{Xian-Lei Sheng$^{1}$, Qing-Bo Yan$^{2}$, Fei Ye$^{2}$,
Qing-Rong Zheng$^{1}$, and Gang Su$^{1}$} \email[Corresponding
author. ]{Email: gsu@gucas.ac.cn} \affiliation{$^1$College of
Physical Sciences, Graduate University of Chinese Academy of
Sciences, P. O. Box 4588, Beijing 100049,
China\\
$^2$College of Materials Science and Opto-Electronic Technology,
Graduate University of Chinese Academy of Sciences, P. O. Box 4588,
Beijing 100049, China}

\begin{abstract}
A structurally stable crystalline carbon allotrope is predicted by
means of the first principles calculations. This allotrope can be
derived by substituting each atom in diamond with a carbon
tetrahedron, and possesses the same space group $Fd\bar{3}m$ as
diamond, which is thus coined as T-carbon. The calculations on
geometrical, vibrational and electronic properties reveal that
T-carbon, with a considerable structural stability and a much lower
density 1.50 g/cm$^3$, is a semiconductor with a direct band gap
about $3.0$ eV, and has a Vickers hardness 61.1 GPa lower than
diamond but comparable with cubic boron nitride. Such a form of
carbon, once obtained, would have wide applications in
photocatalysis, adsoption, hydrogen storage and aerospace materials.
\end{abstract}

\maketitle

Carbon is the element that plays a fundamental role for life on
Earth. As it can form $sp^3$-, $sp^2$- and $sp$-hybridized chemical
bonds, carbon has strong ability to bind itself with other elements
to generate countless organic compounds with chemical and biological
diversity, resulting in the present colorful world. In nature,
elemental carbon has three best-known allotropes, say, graphite,
diamond and amorphous carbon. These forms have been long studied,
and their chemical and physical properties are well disclosed
nowadays. Since 1980s, people have great interest in synthesizing
new allotropes of elemental carbon. The most successful examples
include fullerenes \cite{smalley1985}, carbon nanotubes
\cite{ijima1991}, and graphene \cite{geim2004}. These synthesized
allotropes give rise to enormous scientific and technological
impacts in relevant areas of chemistry, physics, materials and
information sciences, leading to many derivatives, devices and
products. With advances of synthetic tools, a variety of elusive
carbon allotropes such as one-dimensional $sp$-carbyne,
two-dimensional $sp$-$sp^2$-graphyne, and three-dimensional (3D)
$sp$-$sp^3$-yne-diamond, were also obtained or predicted (for review
see e.g. Refs. [\onlinecite{hirsch2010,diederich 2010}]). Recently,
an allotrope of carbon has been obtained by compressing graphite
with pressure over $17$ GPa, whose hardness is even larger than
diamond \cite{mao2003}, while its structure is unknown so far.
Several carbon crystalline phases, e.g., monoclinic M-carbon
\cite{Mcarbon} and bct C$_4$ carbon \cite{bctc4}, were therefore
proposed to simulate this synthesized phase. Moreover, there are
still other experimental and theoretical efforts paid on carbon
(e.g. \cite{anton2010,wangjt2010}). It appears that we might enter
into the era of carbon allotropes \cite{hirsch2010}.

In this Letter, by means of the first principles calculations we
predict a new allotrope of elemental carbon that has very intriguing
physical properties. This structure can be obtained by substituting
each atom in diamond by a carbon tetrahedron, forming a 3D cubic
crystalline carbon that possesses the same space group $Fd\bar{3}m$
as diamond. This form is thus dubbed as the T-carbon. The calculated
results show that T-carbon, kinetically stable and with a much lower
density 1.50 g/cm$^3$, has a Vickers hardness 61.1 GPa smaller than
diamond (93.7 GPa) but comparable with the cubic boron nitride.
Electronic structures reveal that it is a semiconductor with a
direct band gap around 3.0 eV. The possibility of hydrogen storage
in T-carbon is also examined.

\begin{figure}[htbp]
\centerline{\includegraphics[width=7.5cm]{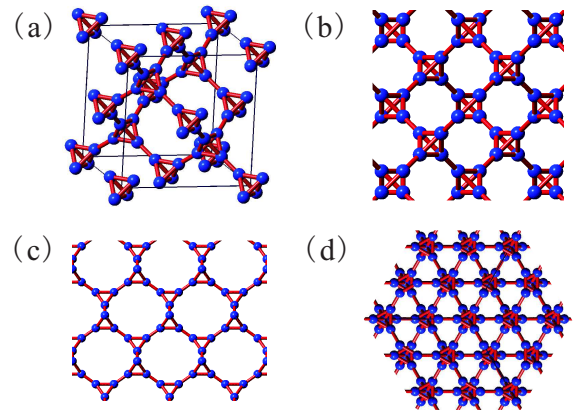}}
\caption[]{(Color online) Schematic depiction of the structure of
T-carbon. (a) The cubic crystalline structure of T-carbon that is
obtained by replacing each atom in diamond by a carbon tetrahedron,
where the lattice constant is $7.52$ {\AA}; (b)-(d) are views from
[100], [110] and [111] directions of T-carbon, respectively.}
  \label{fig:structure}
\end{figure}

Most calculations were performed within the density-functional
theory \cite{dft1,dft2} as implemented within the Vienna \textit{ab
initio} simulation package (VASP) \cite{vasp1,vasp2} with the
projector augmented wave (PAW) method \cite{paw}. Both local density
approximation (LDA) in the form of Perdew-Zunger \cite{ldapz} and
generalized gradient approximation (GGA) developed by Perdew and
Wang \cite{ggapw} were adopted for the exchange correlation
potential. The plane-wave cutoff energy is taken as 400 eV. The
Monkhorst-Pack scheme was used to sample the Brillouin zone
\cite{MPscheme}, and a mesh of 8$\times$8$\times$8 k-point sampling
was used for the calculations. The geometries were optimized when
the remanent Hellmann-Feynman forces on the ions are less than 0.01
eV/{\AA}. The calculations of phonon spectra and electronic
structures were performed using Quantum-ESPRESSO package
\cite{espresso}, where 8$\times$8$\times$8 k-mesh was used.

Inspired by the structural connection between methane and
tetrahedrane, where the latter can be obtained by replacing each
carbon atom in methane with a carbon tetrahedron, we may construct,
likewise, a new allotrope of carbon by substituting each carbon atom
in diamond by a carbon tetrahedron, as indicated in Fig.
\ref{fig:structure}. Luckily, the stability of such a structure was
verified and optimized by the first principles calculations. In this
new structure, each unit cell contains two tetrahedrons with eight
carbon atoms, and the lattice constant is about $7.52$ \AA. The
three unit vectors are $\vec{a}= l/2(1,1,0)$, $\vec{b}= l/2(0,1,1)$
and $\vec{c}= l/2(1,0,1)$. The carbon atoms occupy the Wyckoff
position 32e $(x,x,x)$ with $x=0.0706$. The lattice constant ($l$),
equilibrium density ($\rho$), bond length ($d$), cohesive energy
($E_{coh}$), energy gap ($E_{g}$) between valence and conduction
bands, bulk modulus ($B$) and Vickers hardness ($H_{v}$) at zero
pressure of cubic and hexagonal diamond, graphite, M-carbon, bct
C$_4$, and T-carbon are computed and summarized in Table I for a
comparison. Note that all data presented here, apart from the
sources indicated, were obtained by ourselves using the same method
as the T-carbon, which can thus be compared reasonably at the same
level. It is seen that there exist two distinct bonds with length
$1.502$ {\AA} (intra-tetrahedron) and $1.417$ {\AA}
(inter-tetrahedron) for T-carbon, respectively, both of which are
smaller than those of diamond. There are also two different bond
angles, 60$^\circ$ in tetrahedron, and 144.74$^\circ$ between two
inequivalent bonds. The former is much less than that in diamond
(109.5$^\circ$), implying that a strain exists in T-carbon higher
than that in diamond, which probably leads to the raise of the total
energy. Indeed, our calculations confirm that the cohesive energy
per atom of the T-carbon is $6.573$ eV, around 1 eV/atom smaller
than other forms of carbon at the GGA level, suggesting that this
structure might be less thermodynamical stable against those
allotropes at zero pressure. The remarkable feature is that the
equilibrium density ($\rho$) of T-carbon is the smallest (1.50
g/cm$^3$) among diamond, graphite, M-carbon and bct C$_4$. This is
conceivable, because T-carbon, if compared with other forms of
carbon, has large interspaces between carbon atoms along certain
directions (Fig. 1), and the volume per atom is about twice larger
than those of diamond, M-carbon and bct C$_4$, or 1.5 times larger
than graphite, while its bulk modulus is only 36.4$\%$ of cubic
diamond. Owing to the large interspaces between atoms, the hydrogen
storage in T-carbon may be expected. Our computations show that it
is indeed possible to store hydrogen molecules into the T-carbon,
and the gravimetric and volumetric hydrogen capacities are
calculated to be around 7.7 wt $\%$ and 0.12 Kg H$_2/l$,
respectively.

\begin{widetext}
\begin{table*}[htbp]
\caption{The lattice constant ($l$), equilibrium density ($\rho$),
bond length ($d$), cohesive energy ($E_{coh}$), energy gap
($E_{g}$), bulk modulus ($B$) and Vickers hardness ($H_{v}$) at zero
pressure of cubic diamond (c-diamond), hexagonal diamond
(h-diamond), graphite, M-carbon, bct C$_4$, and T-carbon.}
\begin{tabular}{cccccccccc}
\hline\hline
Structure & Method & $l$ ({\AA})& $\rho$ (g/cm$^3$) & $d$ ({\AA})& $E_{coh}$ (eV/atom) &$E_{g}$ (eV) & $B$ ($10^{2}$ GPa)& $H_{v}$ (GPa)\\
\hline
c-diamond & GGA & 3.566&3.52& 1.544& 7.761&4.16&4.64&93.7\\
c-diamond &LDA & 3.529&3.63 &1.529& 8.859&4.22&5.05&99.0\\
c-diamond &Exp \cite{kittel,andri} &3.567&3.52&1.54&7.37 &5.45 &4.43&96$\pm$5\\
h-diamond &GGA&2.506,4.169 &3.52 &1.539,1.561& 7.732&3.13&4.30&92.8\\
Graphite &Exp \cite{furth}&2.46,(c=2.7a-2.73a) &2.27-2.28&1.42 &7.374 &-&2.86-3.19&\\
Graphite &GGA&2.462,6.849 &2.22 &1.422&7.891&-&2.94\\
M-Carbon&GGA &9.09,2.50,4.10& 3.45&1.488-1.607&7.636&3.56&4.20&91.5\\
Bct C$_{4}$ &GGA& 4.33,2.48 &3.35&1.507,1.561 &7.533&2.47&4.09&92.2\\
T-carbon &GGA &7.52&1.50&1.502,1.417  &6.573&2.25&1.69&61.1\\
T-carbon & LDA &7.45&1.54 &1.488,1.404  &7.503&2.22&1.75&63.7\\
\hline\hline
\end{tabular}

\end{table*}
\end{widetext}

Similar to c-diamond, T-carbon has also three elastic constants,
$C_{11}$, $C_{12}$ and $C_{44}$, which are calculated with GGA to be
225 GPa, 141 GPa and 89 GPa, respectively. With the Voigt-Reuss-Hill
approximation \cite{VGH}, we can acquire the bulk and shear modulus
of T-carbon, say 169 GPa and 70 GPa, respectively. The Vickers
hardness $H_{v}$ of T-carbon is 61.1 GPa that is obtained from an
empirical formula \cite{hardness}
$H(GPa)=350[(N_e^{2/3})e^{-1.191f_i}]/d^{2.5}$, where $N_e$ is the
electron density of valence electrons per {\AA}$^3$, $d$ is the bond
length, and $f_i$ is the ionicity of the chemical bond in a crystal
scaled by Phillips \cite{phillips}. Following the same settings, the
Vickers hardness of diamond is obtained to be 93.7 GPa (in
comparison to the experimental value $96 \pm 5$ GPa). In this sense,
T-carbon is around 1/3 softer than diamond, and is comparable with
the c-BN (whose $H_{v}$ is $64$ GPa \cite{hardness}). These above
features together with a much lower density suggest that the
T-carbon, once synthesized, might have broad applications.

\begin{figure}[htbp]
\centerline{\includegraphics[width=8cm]{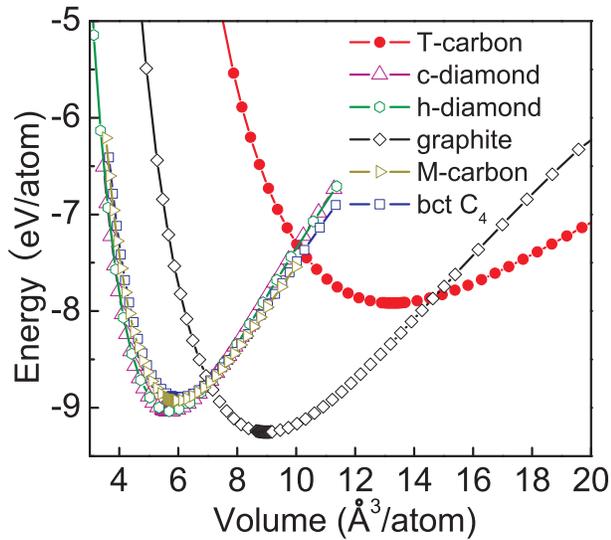}} \caption[]{(Color
online) The total energy per atom as a function of volume per atom
for T-carbon, cubic diamond (c-diamond), hexagonal diamond
(h-diamond), graphite, M-carbon and bct $C_4$ allotropes. }
\label{fig:enthalpy}
\end{figure}

Fig. \ref{fig:enthalpy} plots the total energy per atom against the
volume per atom for the T-carbon, hexagonal diamond, cubic diamond,
graphite, M-carbon and bct C$_4$ allotropes. It can be observed that
the total energy of T-carbon as a function of volume per atom has a
single minimum, showing that the geometrical structure would be
stable. For a comparison, we also calculate the total energies per
atom for other allotropes of carbon, and observe that the minimum
total energy per atom of graphite is the lowest, and the other
allotropes bear higher minimum total energies per atom, while that
of T-carbon is the highest, say -7.92 eV/atom, suggesting that
T-carbon could be a thermodynamically metastable phase against
diamond, graphite, M-carbon and bct-C$_4$ allotropes.

\begin{figure}[htbp]
\centerline{\includegraphics[width=8cm]{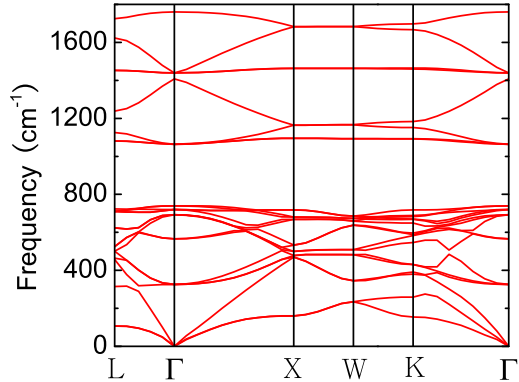}}
\caption[]{\label{fig:phonon} (Color online) Phonon band structure
of T-carbon.}
\end{figure}

To confirm further the mechanical stability, the phonon spectra of
T-carbon are calculated, as shown in Fig. \ref{fig:phonon}. No
imaginary phonon modes in T-carbon are found, implying again the
kinetical stability of T-carbon. The highest phonon frequency of
T-carbon is about 1760 cm$^{-1}$, larger than that of diamond (1303
cm$^{-1}$). It is instructive to note that there exists a wide
direct band gap (325 cm$^{-1}$) of phonons between the frequency 739
- 1064 cm$^{-1}$ at $\Gamma$ point. \emph{}

\begin{figure}[htbp]
\centerline{\includegraphics[width=8cm]{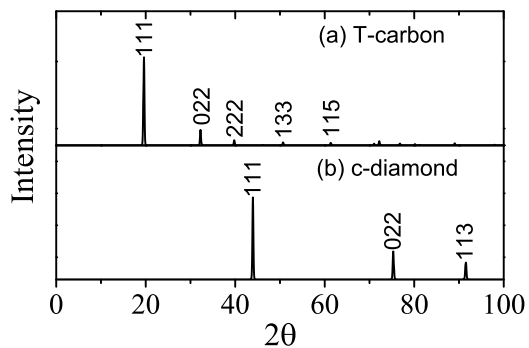}}
\caption[]{\label{fig:xrd} (Color online)  Simulated XRD patterns of
T-carbon and c-diamond. The X-ray wavelength we use is $1.54059$
{\AA}.}
\end{figure}

To provide more information and characters for possible experimental
observation, we also simulate the X-ray diffraction (XRD) spectra of
T-carbon and c-diamond with wavelength $1.54059$ {\AA}. The results
are presented in Fig. \ref{fig:xrd}. Different from diamond where
the peaks of (111) at 2$\theta = 43.94^\circ$, (022) at
$75.31^\circ$ and (113) at $91.51^\circ$ are observed obviously,
only one sharp XRD peak of (111) at $20.44^\circ$ with a strong
intensity is seen for T-carbon, and the peaks of (022) at
$33.68^\circ$, (222) at $39.72^\circ$, (133) at $50.7^\circ$, (115)
at $61.35^\circ$ are negligibly small. This is very similar to
graphite, where a sharp XRD peak of (002) appears with several other
small peaks. These features may be helpful for identifying the
T-carbon in experiments.

\begin{figure}[htbp]
\centerline{\includegraphics[width=9cm]{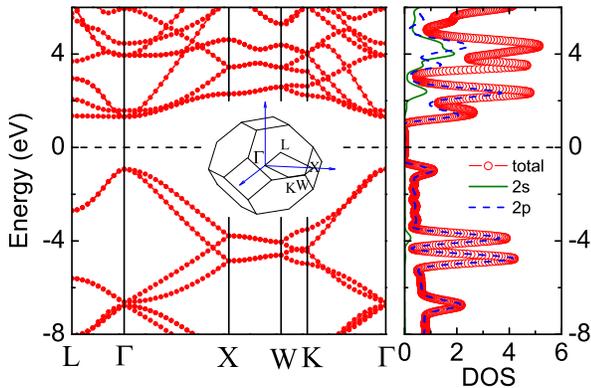}}
\caption[]{\label{fig:energyband} (Color online) Electronic band
structures and density of states (DOS) of T-carbon.}
\end{figure}

Now let us look at the electronic structures and density of states
(DOS) of T-carbon, as given in Fig. \ref{fig:energyband}. One may
see that at the GGA level there is a direct band gap 2.25 eV at
$\Gamma$ point, suggesting that the T-carbon is a direct-band-gap
semiconductor. From the DOS, the band gap is also clearly indicated.
To calculate the band gap more accurately, the B3LYP (Becke,
three-parameter, Lee-Yang-Parr) exchange-correlation functional
\cite{becke, lyp} is used. The band gap of T-carbon is obtained to
be 2.968 eV. We also note that the band gap of T-carbon can be
manipulated by doping, which would be useful for photocatalysis.
With the same method, the band gap of c-diamond is calculated to be
5.799 eV, being comparable with its experimental value 5.45 eV. In
addition, the computations on the projected DOS (PDOS) of each
carbon atom show that the PDOS for $2p_x$, $2p_y$ and $2p_z$
orbitals are the same, much larger than that of $2s$ orbitals,
illustrating that the wave functions of valence electrons in the
T-carbon are recombined, leading to extreme anisotropic
$sp^3$-hybridized bonds. Particularly, the charge population
analysis reveals that the electron density around the
inter-tetrahedron bonds is much higher than that on the
intra-tetrahedron bonds, showing that the inter-tetrahedron bonds
are relatively strong, which are also consistent with their short
bond length, and can balance the strain of the carbon tetrahedron
cage, thereby making the whole structure stable.

In summary, we have predicted a novel carbon allotrope dubbed as
T-carbon that has the same space group $Fd\bar{3}m$ as diamond by
means of the first-principles calculations. This new structure can
be derived by substituting each atom in diamond with a carbon
tetrahedron, and its structural stability is checked and optimized
at both LDA and GGA levels. The calculations on the mechanical and
electronic properties of T-carbon show that it is, with a much lower
density and a bulk modulus than diamond, a semiconductor with a
direct band gap about 3.0 eV. The Vickers hardness of T-carbon is
found $35\%$ smaller than cubic diamond but comparable with that of
cubic boron nitride. The XRD pattern is observed to have features
similar to that of graphite. The possibility of hydrogen storage in
T-carbon is also attested. Upon obtained, the amazing T-carbon would
have wide applications in photocatalysis, adsorption, hydrogen
storage, and aerospace materials.

All calculations are completed on the supercomputer NOVASCALE7000 in
Computer Network Information Center (Supercomputing center) of
Chinese Academy of Sciences (CAS) and MagicCube (DAWN5000A) in
Shanghai Supercomputer Center. This work is supported by the
National Science Foundation of China (Grant Nos. 90922033, 10934008,
10974253, and 11004239), and the CAS.

\end{document}